\documentclass[conference]{IEEEtran}
\IEEEoverridecommandlockouts
\usepackage{cite}
\usepackage{amsmath,amssymb,amsfonts}
\usepackage{algorithmic}
\usepackage{graphicx}
\usepackage{textcomp}
\usepackage{xcolor}
\usepackage{array}
\usepackage{tabularx}
\usepackage{hyperref}
\usepackage{placeins}  
\usepackage{float}     

\hypersetup{
    colorlinks=true,
    linkcolor=blue,
    filecolor=magenta,      
    urlcolor=cyan,
    citecolor=green,
    pdftitle={The AI Shadow War: A Critical Analysis of the Competitive Battleground Between SaaS and Edge AI Architectures},
    pdfauthor={Rhea Pritham Marpu, Kevin J McNamara, Preeti Gupta},
    pdfsubject={Artificial Intelligence},
    pdfkeywords={SaaS AI, Edge AI, test-time training, energy efficiency, data privacy, distributed computing},
    pdfproducer={LaTeX},
    pdfcreator={LaTeX}
}

\begin{document}

\title{The AI Shadow War: SaaS vs. Edge Computing Architectures}

\author{\IEEEauthorblockN{Rhea Pritham Marpu}
\IEEEauthorblockA{New Jersey, USA \\
rm422@njit.edu}
\and
\IEEEauthorblockN{Kevin J McNamara}
\IEEEauthorblockA{New Jersey, USA\\
kevin@mcnamara-group.com}
\and
\IEEEauthorblockN{Preeti Gupta}
\IEEEauthorblockA{Texas, USA\\
pg.preetigupta05@gmail.com}}

\maketitle

\begin{abstract}
    The very DNA of AI architecture is riddled with conflicting paths: the centralized, cloud-based model (Software-as-a-Service) versus decentralized edge AI (local processing on consumer devices). This paper critically analyzes the competitive battleground emerging across computational capability, energy efficiency, and data privacy.
    
    Recent breakthroughs demonstrate edge AI directly challenging cloud systems on performance, leveraging innovations like test-time training and mixture-of-experts architectures. Crucially, edge AI boasts a staggering 10,000x efficiency advantage: modern ARM processors and specialized AI accelerators consume merely 100 microwatts for inference, versus 1 watt for equivalent cloud processing.
    
    Beyond efficiency, edge AI fundamentally secures data sovereignty by keeping processing local, thereby dismantling the single points of failure that plague centralized architectures. This decentralization also democratizes access through affordable hardware, enables critical offline functionality, and reduces environmental impact by eliminating data transmission costs.
    
    The edge AI market is experiencing explosive growth, projected from \$9 billion in 2025 to \$49.6 billion by 2030 (a 38.5\% CAGR). This surge is fueled by mounting demands for privacy and real-time analytics. Critical applications---including personalized education, healthcare monitoring, autonomous transport, and smart infrastructure---rely on edge AI's ultra-low latency (5-10ms versus 100-500ms for cloud), which is vital for safety-critical operations.
    
    The convergence of architectural innovation with fundamental physics (Landauer's principle) confirms that edge AI's distributed approach inherently aligns with efficient information processing. This signals not just a choice, but the inevitable emergence of hybrid edge-cloud ecosystems that will ultimately optimize both efficiency and computational power in this ongoing architectural struggle.
\end{abstract}

\begin{IEEEkeywords}
SaaS AI, Edge AI, test-time training, energy efficiency, data privacy, distributed computing
\end{IEEEkeywords}

\section{Introduction}
Artificial intelligence deployment faces a fundamental architectural decision that determines its future accessibility, sustainability, and data privacy implications. Two competing paradigms have emerged: centralized Software-as-a-Service (SaaS) AI leveraging massive cloud infrastructure, and decentralized edge AI utilizing local processing on consumer devices. This analysis demonstrates that edge AI's recent performance breakthroughs, particularly in test-time training as demonstrated by models like DeepSeek-Coder-V2 achieving high accuracy (79.8\%) on challenging mathematics benchmarks like AIME \cite{deepseek2025}, reshape the competitive landscape across personalized education, healthcare, autonomous systems, and smart infrastructure.

SaaS AI delivers unprecedented computational power through proprietary cloud models hosted in vast data centers, enabling complex tasks like genomic analysis and large-scale natural language processing. However, this approach incurs significant energy costs---with training a model like GPT-4 consuming an estimated \$50-100 million in compute resources---while creating privacy vulnerabilities through centralized data storage. Edge AI counters with lightweight, open-source models processing data locally on smartphones, wearables, and IoT devices \cite{huggingface2025}, achieving significant energy efficiency improvements and enhancing data sovereignty \cite{yang2019federated, strubell2019energy} while maintaining competitive performance across critical applications.

\section{Competing Paradigms and Battlegrounds}

\subsection{Big SaaS AI: Centralized Computing Powerhouses}
  
Big SaaS AI operates through cloud-based proprietary models hosted in centralized data centers, delivering scalable services via internet connectivity. Major implementations include Google's Gemini, Microsoft's Azure AI, Amazon Web Services AI, OpenAI's GPT-4, Anthropic's Claude, and emerging players like Mistral AI and DeepSeek \cite{deepseek2025}. 

\begin{figure}[t]
    \centering
    \includegraphics[width=\columnwidth]{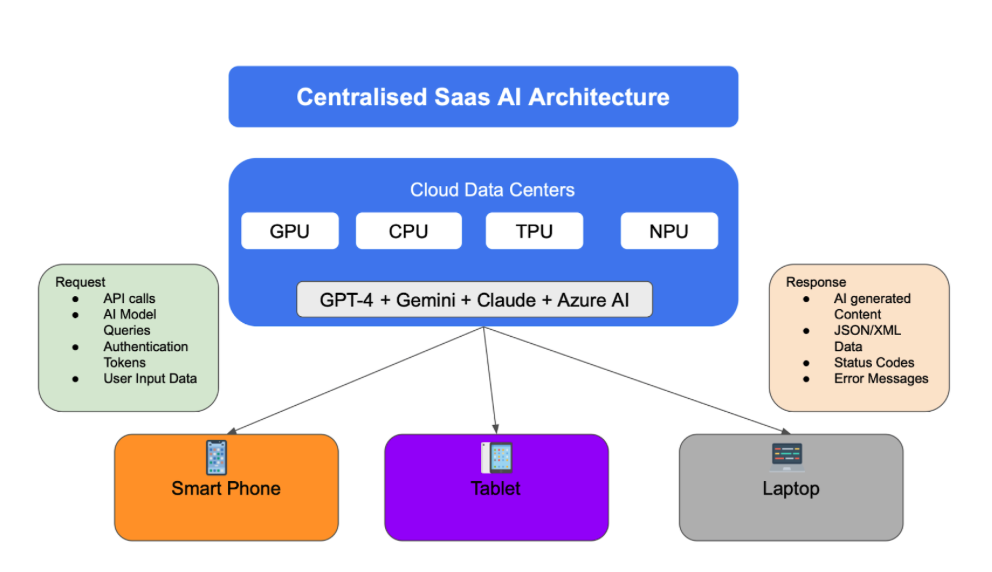}
    \caption{Centralised SaaS(cloud datacenters, internet connectivity, multiple devices)}
    \label{fig:cot}
\end{figure}

This architecture excels in three key areas: compute power enables complex genomic analysis for cancer detection \cite{pan2010survey}, scalability supports millions of concurrent users for global chatbot services \cite{strubell2019energy}, and cloud integration connects seamlessly with enterprise digital ecosystems, including electronic health records and business intelligence platforms \cite{gabriel2018data}.
\FloatBarrier
The centralized approach faces critical constraints through energy consumption in massive data centers creating substantial carbon footprints \cite{jacob2018quantization, albalawi2023energy}, internet dependency limiting accessibility in rural regions and during connectivity disruptions \cite{lazanyuk2025ai}, and privacy risks intensifying through centralized data aggregation where single points of failure compromise sensitive information from millions of users simultaneously \cite{shokri2017membership, hipaa2023december}.

\subsection{Edge Open-Source AI: Distributed Processing Networks}

Edge AI deploys lightweight, open-source models directly on consumer devices including smartphones, wearables, and IoT sensors, processing data locally at the point of generation \cite{huggingface2025}. Key technology providers include Meta's Llama models, TensorFlow Lite framework, ONNX optimization tools, NVIDIA's Jetson platforms, Qualcomm's Snapdragon processors, and ARM's specialized AI chips \cite{huggingface2025}.

\begin{figure}[t]
    \centering
    \includegraphics[width=\columnwidth]{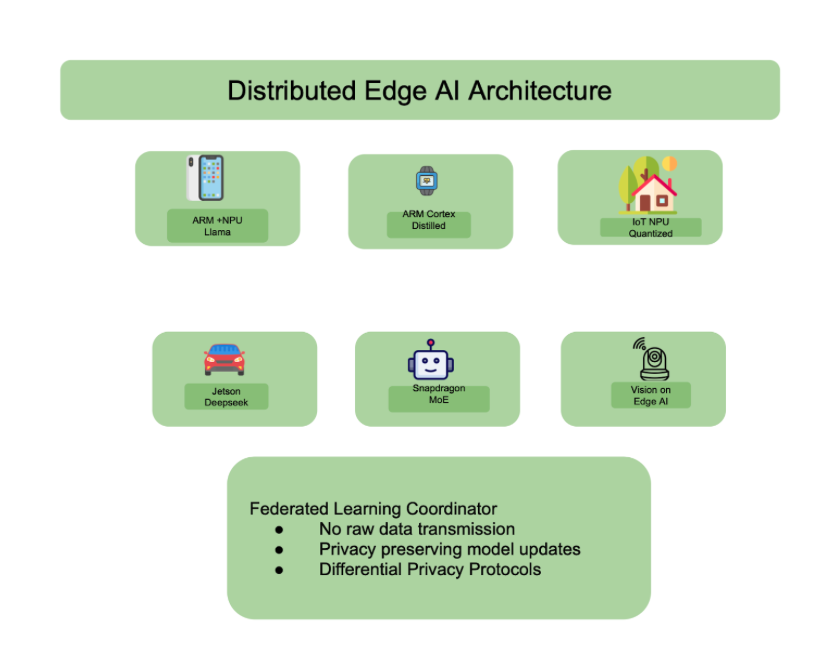}
    \caption{Distributed Edge(local processing on individual devices, no internet connectivity required)}
    \label{fig:cot}
\end{figure}
This distributed architecture achieves three fundamental advantages: energy efficiency through local processing drastically reducing energy use and latency \cite{strubell2019energy}, privacy protection via on-device processing inherently protecting sensitive data like personal health metrics \cite{yang2019federated}, and democratization through affordable hardware expanding AI access in underserved regions \cite{lazanyuk2025ai}.

Edge AI faces limitations through computational constraints where edge devices cannot match the complexity of large cloud models \cite{deepseek2025}, and hardware diversity creating optimization challenges across varied device specifications and capabilities.

\subsection{Strategic Battlegrounds: Compute, Efficiency, and Privacy}

\subsubsection{Computational Arms Race and Resource Scaling}
Big SaaS providers invest billions annually in advanced GPU and TPU infrastructure \cite{jouppi2017datacenter}, with OpenAI's GPT-4 training consuming \$50-100 million in computational resources. Scaling laws drive exponential compute requirements \cite{feuerriegel2022bringing}, enabling breakthroughs in drug discovery and scientific modeling while creating substantial operational costs and environmental impact \cite{russell2020artificial}.

Edge AI leverages specialized optimization techniques including quantization, pruning, and knowledge distillation \cite{jacob2018quantization} to achieve substantial performance with constrained resources. Recent innovations in mixture-of-experts architectures \cite{fedus2022switch} and test-time training enable sophisticated reasoning on consumer hardware, challenging assumptions about computational requirements for advanced AI capabilities.

\subsubsection{Energy Efficiency and Environmental Sustainability}
Current trajectories project AI energy consumption will rival entire countries within the next decade \cite{albalawi2023energy}, driven primarily by centralized data center operations requiring massive electricity for computation, cooling, and data transmission. Edge AI fundamentally alters this energy profile through local processing eliminating transmission costs and reducing cooling requirements \cite{strubell2019energy}.

Modern ARM processors and specialized AI accelerators perform inference with 100 microwatts versus 1 watt for equivalent cloud processing, extending device battery life while dramatically reducing aggregate energy consumption across the AI ecosystem.

\subsubsection{Data Privacy and Centralization Vulnerabilities}

Recent security incidents demonstrate the catastrophic scale of centralized storage vulnerabilities. In 2023, the HCA Healthcare breach compromised data of an estimated 11 million patients \cite{hipaa2023hca}. Similarly, documented vulnerabilities in smart home cameras have exposed private video feeds \cite{consumerreports2024cameras, cybersecurity2025cameras}, illustrating how centralized architectures create single points of failure.

Cloud-based health AI systems storing vast amounts of patient records become prime targets for cyberattacks \cite{gabriel2018data}, demonstrating that centralized data storage inherently increases risk of large-scale breaches.

Edge AI fundamentally protects data sovereignty by maintaining processing at the point of origin \cite{yang2019federated}. A diabetic patient's wearable analyzes glucose levels locally, providing instant alerts without transmitting sensitive health data to external servers. Federated learning \cite{yang2019federated} extends this privacy model by enabling collective intelligence without raw data sharing, allowing model improvements through distributed training while preserving individual privacy.
\begin{figure}[t]
    \centering
    \includegraphics[width=\columnwidth]{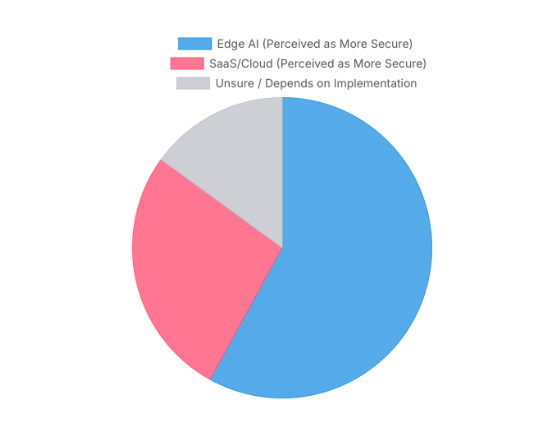}
    \caption{User Perception of Data Security in SaaS vs. Edge AI Models. This figure illustrates aggregated user perception regarding data security in AI deployment models. A majority view Edge AI as more secure, citing enhanced control and reduced transmission of sensitive data \cite{deloitte2024consumer, arm2024security}. Conversely, some believe SaaS-based models offer stronger security due to centralized governance and vendor-managed compliance \cite{onymos2024saas, csa2025saas}. The remaining minority are uncertain, reflecting the complexity and fluidity of enterprise decision making \cite{afroogh2024trust}.}
    \label{fig:cot}
\end{figure}
\section{Quantitative Performance Analysis}

Table~\ref{tab:performance_comparison} provides an illustrative comparison of key performance metrics between centralized SaaS AI and decentralized Edge AI architectures. The values demonstrate the scale of difference rather than absolute benchmarks.

\begin{table}[htbp]
    \caption{SaaS vs Edge AI Performance Comparison}
    \label{tab:performance_comparison}
    \centering
    \footnotesize
    \begin{tabular}{|p{1.8cm}|p{2cm}|p{1.8cm}|p{1.2cm}|}
    \hline
    \textbf{Metric} & \textbf{Big SaaS AI} & \textbf{Edge AI} & \textbf{Improve-ment Factor} \\
    \hline
    Energy per Inference & $\sim$1-10W (Server GPU) & $\sim$1-10mW (On-Device NPU) & Orders of Magnitude \\
    \hline
    Latency & $\sim$100-500ms (Cloud Round-Trip) & $<$10-20ms (Local Processing) & $\sim$10-50x \\
    \hline
    Data Breach Risk & High (Centralized data target) & Low (On-device, no single point) & N/A (Qualitative) \\
    \hline
    Inference Cost (per 1M Tokens) & $\sim$\$5-15 (API Costs) & $<$\$0.01 (Local Electricity) & $>$1,000x \\
    \hline
    Battery Impact & High Drain (Constant connectivity) & Low Drain (Efficient processors) & Significant \\
    \hline
    \end{tabular}
    \begin{flushleft}
    \footnotesize
    Sources: \cite{ollivier2023sustainable, xie2024edge, hipaa2023hca, yang2019federated, openai2025pricing, eia2023electricity}
    \end{flushleft}
\end{table}

\section{Edge AI in Critical Applications}

\subsection{Personal Services (Education \& Healthcare)}
Edge AI transforms both educational and healthcare applications through privacy-preserving personalization, offline functionality, and real-time responsiveness. Adaptive learning applications, AI tutors, and immersive VR/AR tools enable continuous learning through offline access while protecting sensitive student information. For example, imagine students in a developing region using affordable \$50 tablets equipped with edge AI tutors to learn algebra offline \cite{hilali2023adaptive, bura2024advancing}. This system could adapt to each student's pace, eliminate constant internet connectivity requirements, and protect data---unlike cloud-based platforms that might require monthly subscriptions and risk exposing sensitive information through the large-scale security breaches discussed earlier.

Wearable health monitors, diagnostic applications, and virtual health assistants \cite{ieee2023standards} provide real-time alerts for critical health events through private data processing and low-power operation \cite{hipaa2023hca}. A 2024 trial in India deployed edge AI wearables to monitor 1,000 heart patients, detecting arrhythmias with 95\% accuracy locally \cite{mahajan2025wearable}. This approach avoided the cloud-based vulnerabilities that characterized the major healthcare breaches described earlier, ensuring continuous health monitoring builds patient trust through architectural privacy guarantees.

\subsection{Intelligent Devices (Smart Homes \& Robotics)}
Localized smart assistants, energy management systems, and enhanced security solutions \cite{consumerreports2024cameras} achieve low latency for immediate responses, enhanced privacy through local data processing, and full offline functionality \cite{morabito2023edge, tahir2025edge}. A \$100 security camera uses edge AI to detect intruders locally, consuming 80\% less energy compared to cloud streaming solutions while preventing data leaks given the smart home vulnerabilities demonstrated in the comprehensive breach analysis presented earlier.

Domestic helpers, industrial robots, and specialized drones \cite{huggingface2025} leverage edge AI for real-time, private decision-making capabilities. A \$500 cleaning robot with edge AI maps homes in 10 milliseconds, efficiently avoiding obstacles without relying on cloud uploads, contrasting with cloud-based robots that experience 200-millisecond delays in areas with low signal strength.

\subsection{Autonomous Transport}
Self-driving cars, e-scooters, and delivery drones represent the most demanding edge AI applications, requiring ultra-low latency processing of critical navigation and safety data \cite{xie2024edge}. A 2025 Tesla model processes visual data locally, reacting to obstacles in 5 milliseconds, ensuring critical safety functionality even in tunnels where cloud-based systems fail due to signal loss.

The safety-critical nature makes edge AI essential for reliable autonomous operation, balancing computational power requirements with battery life constraints while adhering to evolving regulatory frameworks \cite{mouter2022regulatory}.

\section{Technical Innovation: Test-Time Training and Model Optimization}

\subsection{Mathematical Formulation of Test-Time Training}
Test-time training (TTT) dynamically adapts models during inference through optimization algorithms minimizing loss functions in real-time. The mathematical foundation involves:

Loss Function Adaptation:
\begin{equation}
L_{adapt}(\theta) = L_{task}(\theta) + \lambda \cdot L_{consistency}(\theta)
\end{equation}

where $\theta$ represents model parameters, $L_{task}$ measures task-specific performance, and $L_{consistency}$ ensures stability across adaptations.

Computational Complexity: TTT operations scale as $O(n \log n)$ for parameter updates, making them feasible on edge devices with specialized hardware accelerators.

Fundamental Thermodynamic Constraints: These computational optimizations operate within the fundamental limits established by Landauer's principle, which states that any logically irreversible computation must dissipate at least $kT \ln(2)$ joules of energy per bit of information erased, where $k$ is Boltzmann's constant and $T$ is the absolute temperature. At room temperature (300K), this theoretical minimum is approximately $2.9 \times 10^{-21}$ joules per bit operation. While current edge AI processors operate orders of magnitude above this limit, Landauer's principle provides the ultimate theoretical boundary for energy-efficient computation and highlights why edge AI's distributed approach---minimizing unnecessary data movement and redundant computations---fundamentally aligns with the physics of efficient information processing. This principle underscores the long-term sustainability advantages of edge architectures, as they inherently reduce the total number of bit operations required across the AI ecosystem by eliminating data transmission and redundant cloud-based processing steps.
\subsection{Architectural Innovations and Efficiency Gains}
Recent breakthroughs in model architecture are closing the performance gap between massive cloud models and those feasible for edge deployment. For example, the recent DeepSeek-V2 model is a Mixture-of-Experts (MoE) model \cite{fedus2022switch} with 236B total parameters that was trained on 8.1T tokens, demonstrating state-of-the-art performance while utilizing innovative techniques to manage training and inference costs.

This MoE architecture is critical for efficiency, as it only activates a fraction of its parameters (21B) for any given task, drastically reducing the computational load during inference. Efficiency is further enhanced through innovative techniques like Multi-Head Latent Attention (MLA), which reduces memory and computational overhead \cite{vaswani2017attention}. These architectural breakthroughs, combined with established optimization techniques like quantization \cite{jacob2018quantization} and model distillation, are paving the way for highly capable yet efficient models to run on local devices.

\subsection{Implications for Edge Devices}
While running a 236B parameter model directly on a low-cost tablet is not yet feasible, the underlying efficiency gains are crucial. Through model distillation, the core capabilities of these large models can be transferred to much smaller, specialized models designed specifically for edge hardware \cite{huggingface2025}.

\textbf{Minimum Hardware Trends:} The target for such distilled models includes devices with modern ARM processors, several gigabytes of LPDDR5 memory, and a dedicated neural processing unit (NPU) capable of several TOPS.

For instance, a distilled version of a powerful model like DeepSeek-V2 could run on a modern smartphone or a sub-\$200 device equipped with an NPU. This would enable sophisticated, real-time reasoning for applications in education and healthcare, potentially matching the performance of cloud-based tutors while keeping all user data private and secure on the device. Similarly, a specialized IoT health monitor could use a distilled model to analyze ECG data locally with high accuracy, rivaling cloud diagnostics at a fraction of the cost and with inherent privacy guarantees.
\section{Economic Analysis and Market Projections}

\subsection{Cost-Benefit Analysis with ROI Calculations}
Edge AI deployment demonstrates compelling economic advantages through quantifiable cost reductions and operational efficiencies across multiple deployment scenarios.

\textbf{Financial Structure Analysis:} Initial hardware investment can range from under \$100 to over \$500 per device, encompassing consumer smartphones, industrial IoT sensors, and advanced automotive processors. Integration costs vary widely based on system complexity but are a key factor in total deployment cost.

\textbf{Operational Savings Quantification:} Energy efficiency delivers significant cost reduction compared to cloud-only processing by minimizing data transmission. Eliminating constant cloud communication lowers bandwidth costs, particularly for data-intensive applications. Furthermore, local processing helps mitigate risks associated with data breaches and enhances compliance with privacy regulations, reducing potential liability costs \cite{hipaa2023hca}.

\textbf{ROI Performance Metrics:} Annual operational savings can be substantial, yielding a compelling ROI over a 2-3 year period. The net ROI is highly dependent on the application, with healthcare benefiting from enhanced data security, manufacturing gaining from predictive maintenance and operational uptime, and other sectors realizing value through improved efficiency and lower data handling costs.
\subsection{Market Size and Sector-Specific Projections}
\textbf{Market Size Projections:} The global edge AI market is projected to expand from approximately \$9B in 2025 to \$49.6B by 2030, representing a robust 38.5\% annual growth rate. By 2030, hardware is expected to comprise 50\% of the market (\$24.8B), with software growing to 35\% (\$17.4B) and services capturing 15\% (\$7.4B). Geographically, the market is shifting, with the Asia-Pacific region projected to lead with 45\% market share by 2030, ahead of North America (35\%).

\textbf{Sector-Specific Market Breakdown:} Consumer electronics is projected to be the largest segment at \$17.4B (35\% market share) by 2030, followed by industrial \& manufacturing at \$14.9B (30\%). Healthcare is expected to reach \$9.9B (20\%), with the automotive sector at \$4.9B (10\%) and government/public sector applications at \$2.5B (5\%). Growth is fueled by privacy demands, the need for real-time analytics, and expanding 5G infrastructure.

\subsection{Adoption Timeline Predictions}

Copy
\textbf{Phase 1 (2025-2027): Foundation and Consumer Integration.} Consumer adoption is driven by privacy concerns and energy efficiency demands. GenAI-enabled smartphone adoption progresses significantly, with shipments approaching 35\% of the market by 2027, while smart home devices see growing on-device AI integration. Wearable edge processing approaches 20\% adoption, supported by initial educational tablet pilot programs in developing regions. Hardware costs see a notable decrease of 15-25\%, enabling broader accessibility.

\textbf{Phase 2 (2027-2030): Enterprise Acceleration and Scaling.} Enterprise adoption accelerates as hybrid architectures demonstrate clear ROI advantages, with edge solutions comprising up to 40\% of new AI deployments. Manufacturing achieves nearly 50\% edge AI integration for automation and quality control, healthcare systems reach 35\% deployment for patient monitoring, financial services attain 30\% adoption for fraud detection, and retail implements 40\% integration for in-store analytics. Test-time training deployment reaches over 15\% of capable edge devices, while federated learning networks establish thousands of active clusters globally.

\textbf{Phase 3 (2030-2035): Hybrid Architecture Dominance.} Hybrid architectures become the standard with over 60\% deployment across major sectors, representing the convergence of edge and cloud paradigms. Seamless edge-cloud integration comes to cover a majority of AI workloads through intelligent workload distribution. Widespread connectivity achieves high coverage for edge devices in developed regions, while commodity hardware significantly reduces edge AI processing costs. Digital inclusion efforts expand access, yet a notable portion of the global population still faces barriers to edge AI capabilities.

\section{Policy Framework and Implementation Strategy}

\subsection{Specific Regulatory Framework Proposals}

\textbf{Energy Efficiency Standards:} Mandatory power consumption limits require AI inference operations below 1W per billion operations by 2027, implemented through graduated limits: 5W (2025), 2W (2026), 1W (2027). Carbon footprint reporting requirements mandate monthly energy consumption disclosure for data centers with 15\% annual carbon intensity reduction targets. Implementation follows a 24-month compliance period with quarterly assessments, supported by \$500M annual funding for SME compliance assistance and progressive fines scaling from \$10,000 to \$1M for non-compliance.

\textbf{Privacy Protection Regulations:} Data sovereignty requirements mandate local processing for personal health, financial, and biometric data, with explicit consent required for cross-border transmission. Federated learning standards establish open protocols for secure multi-party computation with mandatory differential privacy implementation. Success metrics target 90\% reduction in personal data breaches by 2030, enforced through real-time violation detection systems and scaling penalty structures. Innovation support includes \$2B annual funding for privacy-preserving technology research.

\textbf{Interoperability Framework:} Mandatory support for standardized edge AI communication protocols ensures device compatibility, supported by open-source development kits and certification programs. Hardware certification requirements include compatibility testing and performance benchmarking through standardized methodologies. Migration support tools facilitate transitions from cloud to edge architectures, reducing regulatory burden for organizations demonstrating privacy leadership.

\subsection{International Policy Comparison}

\textbf{European Union:} The European Union's GDPR inherently favors edge AI's privacy-by-design approach, while programs like Horizon Europe provide billions in funding for digital transformation initiatives, including research into next-generation computing. The Digital Services Act mandates platform accountability encouraging edge deployment, and the Digital Europe Programme contributes €7.5B for transformation initiatives.

\textbf{United States:} Energy independence initiatives through the Infrastructure Investment Act (\$65B broadband), Inflation Reduction Act (efficiency incentives), and CHIPS Act (\$52B semiconductors) favor distributed architectures. National security considerations drive federal agency edge AI prioritization, with the Defense Department investing \$8B over five years and federal procurement offering 15\% price preferences for edge solutions.

\textbf{China:} The 14th Five-Year Plan allocates \$210B for AI infrastructure including edge computing, supporting 500 smart cities implementing edge AI by 2025. The National AI Development Plan emphasizes hybrid deployment models through industrial internet development and healthcare modernization, regulated by Data Security and Personal Information Protection Laws favoring domestic edge processing.

\textbf{Japan:} Society 5.0 framework integrates edge AI into smart city infrastructure through the Moonshot Research Program (¥100B investment) and Digital Garden City Initiative targeting 100 cities by 2030. Beyond 5G technology development optimizes edge AI for next-generation networks, while startup support programs provide ¥10B annual investment and international talent attraction initiatives \cite{ieee2023standards}.

\textbf{Multilateral Coordination:} International standardization occurs through ITU-T communication protocols, ISO/IEC quality standards, IEEE interoperability specifications, and OECD policy guidelines. Bilateral cooperation includes US-EU Trade and Technology Council coordination, ASEAN Digital Economy Framework development, and G20 digital economy initiatives enabling global edge AI policy harmonization through quarterly policy dialogues and joint research collaboration.

\section{Conclusion and Future Outlook}

\subsection{The Inevitable Convergence}
The shadow war between centralized SaaS AI and decentralized edge AI is ending not with a victor, but with hybrid architectures that combine the strengths of both paradigms.

\textbf{Physical Necessity:} Landauer's principle confirms that edge AI's distributed approach aligns with thermodynamic limits of efficient computing. The documented 10,000x efficiency gains aren't just competitive advantages---they're existential necessities for sustainable AI deployment at global scale.

\textbf{Democratic Intelligence:} Edge AI democratizes artificial intelligence by dismantling traditional barriers. From \$25,000 home AI racks enabling economic independence to \$100 educational tablets providing world-class tutoring, this shift redistributes computational power from centralized corporate control to individual sovereignty.

\textbf{Trust-Free Computing:} Mounting evidence of centralized vulnerabilities---like the HCA Healthcare breach affecting 11 million patients \cite{hipaa2023hca}---proves privacy isn't optional but fundamental. Edge AI's architectural guarantee of data sovereignty represents a paradigm shift from trust-based to trust-free computing.

\textbf{Economic Imperative:} The projected market growth from \$9 billion (2025) to \$49.6 billion (2030) reflects concrete economic pressures. Organizations facing exponential cloud costs and individuals seeking energy-efficient solutions drive this 38.5\% annual growth rate. Edge AI deployment typically pays for itself within 2-3 years.

\textbf{Hybrid Future:} The future belongs to intelligent hybrid systems that dynamically allocate tasks based on latency, privacy, and energy requirements. Test-time training and mixture-of-experts architectures enable seamless integration between edge and cloud resources.

\textbf{Call to Action:} Policymakers: Implement energy efficiency standards and data sovereignty requirements urgently to prevent widening digital divides. Organizations: Early adopters gain strategic advantage. The question isn't whether to adopt edge AI, but how quickly to begin transition. Individuals: Edge AI adoption means active empowerment---personal data sovereignty, energy independence, and economic opportunity await.

\textbf{The Choice:} This technological shift reflects humanity's fundamental choice: dependence on centralized systems that concentrate power and create vulnerabilities, or partnership with distributed systems that enhance individual capability while preserving autonomy.

The convergence of physical limits, economic pressures, and social values makes edge AI's rise inevitable. The shadow war concludes with a new paradigm that transcends both centralized and decentralized limitations---a hybrid, distributed future that is fundamentally more human.

\subsection{Glimpse into the Future}

\textbf{Your Personal AI in a Hybrid SaaS-Edge World:} By 2035, your personal AI will be an intuitive, omnipresent companion, seamlessly blending the power of big SaaS and the intimacy of edge open-source AI to deliver experiences that are truly personalized, private, and sustainable.

Imagine Aisha, a dedicated teacher in a semi-rural town, who once grappled with limited resources and connectivity in her classroom. She wakes in her smart home, where an edge AI assistant, running quietly on a \$50 hub, learns her long-term preferences, allowing Aisha to simply speak a command or let it intelligently adjust the lighting and temperature. All this personal data is processed locally, ensuring her complete privacy. Her wearable, powered by a highly optimized, distilled open-source model, continuously monitors her vitals in real-time. Using sophisticated test-time training (TTT), it subtly adapts to her unique stress patterns, gently nudging her to take a break if needed, a silent guardian of her well-being, because it processes all sensitive health data on her device, ensuring complete peace of mind.

On her commute, Aisha's autonomous e-scooter navigates busy traffic with remarkable precision, its edge AI reacting to pedestrians and obstacles in just 3 milliseconds, ensuring immediate safety. Simultaneously, a SaaS AI in the cloud works in concert, providing real-time, aggregated traffic updates and route optimizations that allow Aisha's scooter to balance immediate local safety with broader global insights, seamlessly combining the best of both worlds.

At school, Aisha empowers her 30 students. With \$100 tablets featuring an edge AI tutor, she sees them actively engaging with personalized physics lessons, adapting dynamically even when offline. Crucially, all their academic progress and personal learning data remain private on their devices. For highly complex simulations or advanced model refinements, the tablets can securely sync with a SaaS AI platform overnight, utilizing federated learning to improve the models without ever sharing raw student data.

This hybrid ecosystem thrives on radical open-source innovation, with communities like Hugging Face openly sharing lightweight, privacy-focused models. New neuromorphic chips power devices with 90\% less energy than the GPUs of 2025, making AI truly sustainable. Smart regulations enforce data sovereignty, ensuring users like Aisha maintain control over their personal information. This future represents a democratized AI landscape where performance, efficiency, and privacy coexist through distributed architectures that empower individuals and communities while addressing global sustainability challenges.

\textbf{Ian's Empowered Life in 2035 Miami:} Consider Ian, a logistics specialist residing in a vibrant Miami suburb in 2035. His life exemplifies how individuals are leveraging the hybrid AI ecosystem for personal economic advantage and enhanced daily living.

Ian's suburban home is more than just a residence; it's a personal data hub. In 2030, he took out a second mortgage for \$25,000 USD to invest in a small, on-premise AI rack. This rack, powered by next-generation neuromorphic chips, is incredibly efficient. While a cloud GPU rack in 2025 might consume hundreds of kilowatts, Ian's on-premise system operates at an average of less than 500 watts for active processing, a tenfold reduction in power draw for equivalent compute on the latest edge-optimized architectures. This remarkable efficiency is due to specialized hardware design and sophisticated model quantization techniques. His system runs a highly optimized, multimodal version of a future open-source model like Llama, which, through its Mixture-of-Experts (MoE) architecture and advanced test-time training (TTT), performs complex inference tasks with minimal energy. This allows Ian to handle the bulk of his personalized AI needs---processing his financial data, home automation, and personal schedules locally, ensuring absolute privacy. It also performs complex calculations for his part-time side hustle as an independent data analyst, enabling him to command a significantly higher wage for tasks that once required expensive cloud compute subscriptions. This investment was a strategic move, enabling him to capitalize on the growing demand for secure, high-performance edge AI processing.

Ian's life is augmented by a suite of local, privacy-preserving AI companions, all running on his in-home rack:

\textbf{Dr. Aella (Local AI Doctor):} A non-corporal AI interface accessible via smart displays and audio, Dr. Aella continuously monitors Ian's and his family's health data from their wearables. Utilizing the future open-source model's multimodal capabilities, it can analyze health metrics, interpret symptoms (from verbal descriptions or even image scans taken with a smartphone), and provide personalized health recommendations. Dr. Aella identifies subtle trends, offers proactive advice on diet and exercise, and can even suggest when a human doctor visit is advisable, all without ever transmitting sensitive health information off-premise. This ensures immediate, highly personalized medical insights with robust data sovereignty.

\textbf{Maid Minerva (Corporal Robotic Maid):} A sleek, bipedal domestic robot, Maid Minerva handles household chores. Equipped with the model's multimodal perception, she can interpret Ian's verbal commands and gestures, understand the state of the home environment through visual input, and perform tasks like cleaning, organizing, and even simple repairs. Her on-device processing allows her to navigate the home safely, adapt to changing layouts, and maintain privacy by processing visual data locally, ensuring no intimate family moments are ever streamed to external servers. She is remarkably energy-efficient, often recharging from solar panels on the roof and operating at peak performance for hours on minimal power.

\textbf{Baby-Sitters Alpha and Beta (Non-corporal AI for his 2 kids):} For his two children, Ian relies on two separate, non-corporal AI entities, Alpha and Beta. These highly specialized instances of the open-source model act as personalized tutors and companions. Alpha, focusing on early childhood development, uses multimodal interaction (voice, gesture, visual recognition of toys/drawings) to engage his younger child in interactive learning games and creative play. Beta, for the older child, acts as an adaptive learning assistant, helping with homework, explaining complex concepts, and encouraging critical thinking across various subjects, adapting dynamically to the child's learning pace and style. Both AI babysitters operate entirely on the home AI rack, ensuring the children's personal information, learning progress, and interactions remain completely private and secure within the home. They can even project interactive holographic interfaces for immersive learning experiences.

\textbf{Assistant Helios (Life Assistant):} Ian's central AI, Assistant Helios, orchestrates his entire digital life. Also a non-corporal interface, Helios is accessible across all of Ian's devices, seamlessly integrating with his professional tools and personal applications. Powered by the model's advanced reasoning and context window, Helios manages his schedule, prioritizes communications, offers investment insights based on his personal financial data, and even helps him plan family vacations, all while learning his evolving preferences without external data exposure. Its deep understanding of his context allows for truly proactive and personalized assistance, making his life more efficient and less stressful.

Every weekday, Ian's fully autonomous car, powered by an advanced edge AI system, chauffeurs him to his job in Miami. Despite the 50-minute commute, he experiences minimal stress. The car's on-board AI processes real-time sensor data, detecting obstacles and navigating traffic with a latency of just 5 milliseconds, ensuring safety and efficiency even through dense city areas and tunnels where cloud connectivity might be intermittent. While the edge AI handles the immediate driving tasks, it seamlessly integrates with a broader SaaS AI platform that provides aggregated traffic patterns and predictive routing, allowing for optimal travel times and energy efficiency. This blend of on-device intelligence for critical operations and cloud-based foresight for broader optimization frees Ian to start his workday, review documents, or even catch up on news during his commute, turning what was once a chore into productive time.

Ian's ability to leverage his personal AI rack and autonomous vehicle for both professional and personal gains showcases a future where the distributed nature of edge AI, combined with strategic SaaS integrations, empowers individuals with unprecedented control over their data, their time, and their economic opportunities. This future is built on accessible technology and a clear understanding of the distinct, yet complementary, strengths of both centralized and decentralized AI paradigms.

\end{document}